\batchmode

\documentclass[prb,superscriptaddress,showpacs,preprint]{revtex4-1}
\usepackage[T1]{fontenc}
\usepackage[latin9]{inputenc}
\usepackage{color}
\usepackage{array}
\usepackage{prettyref}
\usepackage{textcomp}
\usepackage{multirow}
\usepackage{amsmath}
\usepackage{graphicx}
\usepackage{esint}
\usepackage{subscript}

\makeatletter

\DeclareRobustCommand{\greektext}{%
  \fontencoding{LGR}\selectfont\def\encodingdefault{LGR}}
\DeclareRobustCommand{\textgreek}[1]{\leavevmode{\greektext #1}}
\DeclareFontEncoding{LGR}{}{}
\DeclareTextSymbol{\~}{LGR}{126}
\providecommand{\tabularnewline}{\\}

\@ifundefined{textcolor}{}
{%
 \definecolor{BLACK}{gray}{0}
 \definecolor{WHITE}{gray}{1}
 \definecolor{RED}{rgb}{1,0,0}
 \definecolor{GREEN}{rgb}{0,1,0}
 \definecolor{BLUE}{rgb}{0,0,1}
 \definecolor{CYAN}{cmyk}{1,0,0,0}
 \definecolor{MAGENTA}{cmyk}{0,1,0,0}
 \definecolor{YELLOW}{cmyk}{0,0,1,0}
}

\makeatother

\begin{document}

\preprint{This line only printed with preprint option}

\title{Effect of Ni on Cu precipitation kinetics in $\alpha$\textendash{}Fe
by AKMC study}

\author{Yi Wang}

\email{yiwang_cn@126.com}

\affiliation{School of Materials Science and Engineering, Nanjing University of
Science and Technology, Nanjing 210094, People\textquoteright{}s Republic
of China}

\author{Huai Yu Hou}

\email{hyhou@njust.edu.cn}

\affiliation{School of Materials Science and Engineering, Nanjing University of
Science and Technology, Nanjing 210094, People\textquoteright{}s Republic
of China}

\author{Xiang Bing Liu}

\affiliation{Center of Plant Life Management, Suzhou Nuclear Power Research Institute,
Suzhou 215004, China}

\author{Rong Shan Wang}

\affiliation{Center of Plant Life Management, Suzhou Nuclear Power Research Institute,
Suzhou 215004, China}

\author{Jing Tao Wang}

\email{jtwang@njust.edu.cn}

\affiliation{School of Materials Science and Engineering, Nanjing University of
Science and Technology, Nanjing 210094, People\textquoteright{}s Republic
of China}
\begin{abstract}
The kinetics of coherent Cu rich precipitation in Fe\textendash{}Cu
and Fe\textendash{}Cu\textendash{}Ni alloys during thermal ageing
have been modeled by Atomic Kinetic Monte Carlo method (AKMC). The
AKMC is parameterized by existing \textit{ab-inito} data to treat
vacancy mediated diffusion which is depend on local atomic environment.
A nonlinear semi-empirical time adjusting method is proposed to rescaled
the MC time. The combining AKMC and time adjusting method give a good
agreement with experiments and other simulations, including advancement
factor and the Cu cluster mobility. Simulations of ternary alloys
reveal Ni has a temporal delay effect on Cu precipitation. This effect
is caused by the decreasing of diffusion coefficient of Cu clusters.
And the reduction effect of diffusion coefficient weakens with larger Cu cluster
size. The simulation results can be used to explain the experimental phenomenon
that ternary Fe\textendash{}Cu\textendash{}Ni alloys have higher cluster number density
 than corresponding binary 
alloy during coarsening stage, which is related to cluster mobility.
\end{abstract}


\maketitle

\section{introduction}

Irradiation induced precipitation is believed to be one major reason
for the degradation of mechanical properties of alloys in radiation
environments. In $\alpha$\textendash{}Fe, which is the basics of
ferritic steels, Cu rich precipitation from supersaturated matrix
is greatly accelerated by irradiation. This kind of Cu rich precipitation
is the primary reason of embrittlement for reactor pressure vessel
(RPV) steels at low doses compared to the so called \textquotedblleft{}late
blooming phases\textquotedblright{} of MnNi rich precipitations\cite{CambridgeJournals:8104390}
at high doses. Experiments show third-party elements such as Ni, Mn
and Si etc. also exist in Cu rich precipitations and may have influence
on Cu precipitation kinetics. In those elements, Ni and Mn are the
richest elements in typical RPV steels\cite{A508_2005}. Possibly,
they also have the strongest influence on Cu precipitation. Miller
\textit{et al.}\cite{Miller2003133,Miller2006187} found that a Fe\textendash{}Cu\textendash{}Mn
model steel has the cluster density approximately an order of magnitude
higher than that of Fe\textendash{}Cu steel, and RPV steels with high
nickel content may have retarded precipitation growth as evidenced
by smaller cluster size. Meslin \textit{et al.}\cite{Meslin2010137}
discovered that advancement of precipitation is lower in the presence
of Mn and Ni, suggesting they may delay the copper precipitation.
The common summaries of experiments are higher precipitation number
density or lower advancement is found in ternary alloys within same
ageing time or radiation dose of binary alloy.

To further reveal the mechanism of the influence by third-party elements,
atomic level computer simulation solutions seem to be attractive.
Atomic kinetic Monte Carlo (AKMC) method based on diffusion of point
defects has become an effective research tool on precipitation for
having detailed information on atomic configuration in full time scale
and being convenient to separate different factors. Vincent\textit{
et al.}\cite{Vincent2008154}\textit{ }studied the effect of Mn and
Ni on Cu precipitation during radiation flux, the simulation results
show Mn containing alloy has slightly smaller cluster size, while
Ni seems to have little influence. Bonny \textit{et al.}\cite{doi:10.1080/14786430903299824}\textit{
}applied an artificial neural network (ANN) powered AKMC to study
the precipitation of a ternary Fe\textendash{}Cu\textendash{}Ni alloy,
and found the peak density of clusters increased by about 29\% than
binary alloy. However, due to the time evolution model used by previous
works, the kinetics is not represented in view of real time evolution,
e.g. MC time scale is found incomparable to experiments\cite{doi:10.1080/14786430903299824},
which limited further comparison.

In this study, we focus on the effect of Ni on Cu precipitation kinetics.
To accomplish our aims, an AKMC approach is applied to simulate thermal
ageing of Fe\textendash{}Cu and Fe\textendash{}Cu\textendash{}Ni alloys.
The following content of this paper is divided into three parts. Detail
of the computational methods used in the AKMC is presented in the
first part. The parameterization is given and a nonlinear
time adjusting method based on post-processing of AKMC data is proposed
in order to reflect the kinetics correctly. In the second part, the
simulation results are reported. Firstly, the simulation results of precipitation kinetics
of the Fe\textendash{}Cu binary system at varying temperatures
are used to verify applicability of AKMC parameters and combined time
adjusting method. Then the results of alloys with different Ni content,
aged at same temperature, are compared. In the final part of the paper,
the effect of Ni on Cu precipitation kinetics is discussed.

\section{methods}

In the AKMC simulation, the precipitation process is induced by thermal ageing. Initially, Cu
and Ni substitutional atoms are randomly introduced into $\alpha$\textendash{}Fe
matrix. A single vacancy is randomly introduced into the system to
treat vacancy mediated diffusion.

\subsection{AKMC simulation model}

An AKMC code has been developed at Nanjing
University of Science and Technology, with rigid on-lattice model
(RLM) and Bortz\textendash{} Kalos\textendash{} Lcbmvitz (BKL) algorithm\cite{PhysRevB.12.2000,Bortz197510}.
When the vacancy lies on the first nearest neighbor lattice site of
one atom, the probability of the position exchanging between this
atom and the vacancy is obtained by the Arrhenius equation,
\begin{equation}
\mathit{\Gamma}_{X}=\nu_{X}\mathrm{exp}\left(-\frac{E_{a}}{k_{\mathrm{B}}T}\right),\label{eq:arrhenius}
\end{equation}
where $\nu_{X}$ is the attempt frequency for atom species \textit{X},
which is related to local vibration modes\cite{Vineyard1957121}.
In the present paper, $\nu_{X}$ values for all atom species are taken
as one independent constant of 6\texttimes{}10\textsuperscript{12}
s\textsuperscript{-1}, which is on the same order of Debye frequency.

The time evolution of one KMC step is given by the summation of jump
frequency of every possible exchanging,

\begin{equation}
\mathrm{\delta}t=-\frac{\ln R}{\sum\mathit{\Gamma}_{X}},
\end{equation}
where, \textit{R} is a uniform random number between 0 and 1.

\subsection{Activation energy model}

The activation energy in eq. \prettyref{eq:arrhenius} plays the key role
in diffusion dynamics. Comprehensive description on energy models
with heuristic formulas has been reported in Ref. \onlinecite{Vincent2008154,PhysRevB.76.214102,PSSB:PSSB200945251,PhysRevB.65.094103}.
Recently Vincent \textit{et al.}\cite{Vincent2008387} made a critical
review of these models. It is also possible to directly predict the
activation energy by ANN AKMC\cite{Djurabekova20078,castin:074507,Castin20093002}.
Here, we use the final initial system energy model (FISE) \cite{PSSB:PSSB200945251,PhysRevB.76.214102,Vincent2008154},
which has the same form of the Kang\textendash{}Weinberg decomposition\cite{kang:2824}.
For the situation of atom species \textit{X} exchanging with a vacancy
lying on the first nearest site, the activation energy writes as,
\begin{equation}
E_{a}\doteq\frac{E_{fnl}-E_{ini}}{2}+Q_{X},
\end{equation}
where \textit{E\textsubscript{ini}} and \textit{E\textsubscript{fnl}}
are the initial and final system energies, respectively. \textit{Q\textsubscript{X}}
is the migration energy of atom species \textit{X} in $\alpha$\textendash{}Fe
matrix.

The interaction of atoms is ranged up to second nearest neighbor and
under two-body approximation, thus the system energy at a specific
state is evaluated by summation of the energies of pairwise bonds,
as following,
\begin{equation}
E=\overset{2}{\underset{i=1}{\sum}}\underset{j=1}{\overset{Zi}{\sum}}\varepsilon_{X-A_{j}}^{\left(i\right)}+\overset{2}{\underset{i=1}{\sum}}\underset{j=1}{\overset{Zi}{\sum}}\varepsilon_{\mathrm{V-}B_{j}}^{\left(i\right)}-\varepsilon_{X-\mathrm{V}}^{\left(1\right)},
\end{equation}
where \textit{i} represents the bonds are counted in both first (\textit{i}
=1) and second (\textit{i} =2) nearest neighbor sites. \textit{Z\textsubscript{\textit{i}}}
is the coordination number at each distance, \textit{A\textsubscript{\textit{j}}}
is a neighbor atom of the jumping atom \textit{X} , \textit{B\textsubscript{\textit{j}}}
is a neighbor atom of the vacancy, \textit{$\varepsilon$} is bond
energy. \textit{X}-V bond at first nearest neighbor distance is double
counted, thus is subtracted in the third term.

Basic information on atom interactions is usually obtained by \textit{ab-initio}
calculations in a multi-scaled fashion. Results obtained by Vincent
\textit{et al.}\cite{Vincent2008,Vincent2008154,Vincent2005137} has
been opted in current study. The atomic interactions in Fe\textendash{}Cu\textendash{}Ni\textendash{}Mn\textendash{}Si
system were studied using Projector Augmented-Wave (PAW) and UltraSoft
Pseudo Potential (USPP) in their works. It was found that the USPP
results reached a better agreement with experiments\cite{Vincent2008}.
The values of \textit{Q\textsubscript{\textit{X}}} for Fe, Cu and
Ni are 0.62eV, 0.54eV and 0.68eV respectively\cite{Vincent2008}.
In the present work, we fitted pairwise bond energies using a method
similar to the description in Ref. \onlinecite{Vincent2008154}. The
\textit{ab-initio} data of cohesive energy, mixing energy, binding
energy and vacancy formation energy are expanded by following relations
in RLM,
\begin{equation}
E_{X}^{coh}=4\varepsilon_{X-X}^{\left(1\right)}+3\varepsilon_{X-X}^{\left(2\right)},\label{eq:coh}
\end{equation}
\begin{equation}
E_{X}^{sol}\left(\mathrm{Fe}\right)=-4\varepsilon_{\mathrm{Fe-Fe}}^{\left(1\right)}-3\varepsilon_{\mathrm{Fe-Fe}}^{\left(2\right)}+8\varepsilon_{\mathrm{Fe}-X}^{\left(1\right)}+6\varepsilon_{\mathrm{Fe}-X}^{\left(2\right)}-4\varepsilon_{X-X}^{\left(1\right)}-3\varepsilon_{X-X}^{\left(2\right)},\label{eq:sol}
\end{equation}
\begin{equation}
E_{XY}^{b\left(i\right)}\left(\mathrm{Fe}\right)=-\varepsilon_{\mathrm{Fe-Fe}}^{\left(i\right)}+\varepsilon_{\mathrm{Fe}-X}^{\left(i\right)}+\varepsilon_{\mathrm{Fe}-Y}^{\left(i\right)}-\varepsilon_{X-Y}^{\left(i\right)},\label{eq:bind}
\end{equation}
\begin{equation}
E_{\mathrm{V}}^{for}\left(X\right)=8\varepsilon_{X-\mathrm{V}}^{\left(1\right)}+6\varepsilon_{X-\mathrm{V}}^{\left(2\right)}-4\varepsilon_{X-X}^{\left(1\right)}-3\varepsilon_{X-X}^{\left(2\right)},\label{eq:forv}
\end{equation}

To balance the ratio between bond energies of \textit{X}-\textit{Y}
pair on the first nearest and second nearest distance, extra equations
are needed. Vincent \textit{et al.\cite{Vincent2008154}} assumed
a constant ratio between the second nearest bonds of \textit{X}-\textit{X}
pair and Fe-Fe pair. Similarly, in Ref. \onlinecite{PhysRevB.76.214102},
except the Cu-V bonds, the energy of a second nearest \textit{X-Y}
bond is half of the first nearest \textit{X-Y} bond. Here we added
extra equations by assuming the interfacial energies on \{100\}, \{110\}
and \{111\} planes have negligible difference, as the following equations,
\begin{equation}
E_{XY}^{\mathrm{int}\left\{ 110\right\} }=E_{XY}^{\mathrm{int}\left\{ 100\right\} },\label{eq:int100}
\end{equation}
\begin{equation}
E_{XY}^{\mathrm{int}\left\{ 110\right\} }=E_{XY}^{\mathrm{int}\left\{ 111\right\} },\label{eq:int111}
\end{equation}

The interfacial energies can be expanded as equations,
\begin{equation}
E_{XY}^{\mathrm{int}\left\{ 100\right\} }=-2\varepsilon_{X-X}^{\left(1\right)}-\varepsilon_{X-X}^{\left(2\right)}+4\varepsilon_{X-Y}^{\left(1\right)}+2\varepsilon_{X-Y}^{\left(2\right)}-2\varepsilon_{Y-Y}^{\left(1\right)}-\varepsilon_{Y-Y}^{\left(2\right)},
\end{equation}
\begin{equation}
E_{XY}^{\mathrm{int}\left\{ 110\right\} }=\left[-\varepsilon_{X-X}^{\left(1\right)}-\varepsilon_{X-X}^{\left(2\right)}+2\varepsilon_{X-Y}^{\left(1\right)}+2\varepsilon_{X-Y}^{\left(2\right)}-\varepsilon_{Y-Y}^{\left(1\right)}-\varepsilon_{Y-Y}^{\left(2\right)}\right]\cdot\sqrt{2},
\end{equation}
\begin{equation}
E_{XY}^{\mathrm{int}\left\{ 111\right\} }=\left[-\frac{5}{2}\varepsilon_{X-X}^{\left(1\right)}-3\varepsilon_{X-X}^{\left(2\right)}+5\varepsilon_{X-Y}^{\left(1\right)}+6\varepsilon_{X-Y}^{\left(2\right)}-\frac{5}{2}\varepsilon_{Y-Y}^{\left(1\right)}-3\varepsilon_{Y-Y}^{\left(2\right)}\right]/\sqrt{3},
\end{equation}

Practically, our fitting is divided into two steps. In the first step,
equations \prettyref{eq:coh}\prettyref{eq:sol}\prettyref{eq:bind}\prettyref{eq:int100}\prettyref{eq:int111}
are used to get the interactions between atoms. Then, in the second
step, equations \prettyref{eq:bind} and \prettyref{eq:forv} are
used to get the interactions between atoms and vacancy, where the
interactions between atoms obtained by first step are considered to
be known. In both steps, the equation sets are over-determined, the
Moore Penrose generalized inverse matrix method was used to get the
least square solution.

\begin{table}

\caption{Fitting result of pairwise bonds of atomic interactions\label{tab:Fitting-result}}
\begin{ruledtabular}

\begin{tabular}{ccc}
\multirow{2}{*}{\textcolor{black}{Bond }\textit{\textcolor{black}{X}}\textcolor{black}{-}\textit{\textcolor{black}{Y}}} &
\multicolumn{2}{c}{$\varepsilon_{X-Y}^{\left(i\right)}$ (eV)}\tabularnewline
\cline{2-3} 
 & \textit{i}=1 &
\textit{i}=2\tabularnewline
\hline 
Fe-Fe &
-0.6856 &
-0.5125\tabularnewline
Fe-Cu &
-0.5696 &
-0.4429\tabularnewline
Fe-Ni &
-0.7099 &
-0.5201\tabularnewline
Cu-Cu &
-0.5610 &
-0.4153\tabularnewline
Cu-Ni &
-0.6600 &
-0.4764\tabularnewline
Ni-Ni &
-0.6903 &
-0.5263\tabularnewline
Fe-V &
-0.2117 &
-0.1035\tabularnewline
Cu-V &
-0.1853 &
-0.1611\tabularnewline
Ni-V &
-0.2601 &
-0.2767\tabularnewline
\end{tabular}

\end{ruledtabular}
\end{table}

The fitted pairwise bond energies $\varepsilon_{X-Y}^{\left(i\right)}$
are given in Table \ref{tab:Fitting-result}. It's worthwhile to notice
that because the fitting equation sets are over-determined, not all
equations are exactly equal on both sides at the end. Actually, only
cohesive energies of Fe, Cu, Ni were exactly fitted, the values are
-4.28eV, -3.49eV and -4.34eV respectively. Comparison of other energies
is presented in Table \ref{tab:Comparison-of-energies}. The binding
energies of Cu-Cu and Cu-V are obviously smaller than the \textit{ab-intio}
data. Though, mixing energy of Cu in $\alpha$\textendash{}Fe matrix
was fitted well. We think it is because Cu-\textit{X} pairs have strong
many-body contribution that cannot be reproduced by pairwise bonds.
As is shown in Ref. \onlinecite{Soisson2007}, cluster expansion can
get better fitting. It is also need to mention that for Cu and Ni,
the reference structure for cohesive energies and mixing energies
in Vincent's original work are fcc\cite{Vincent2008,Vincent200688}.
Energy difference between fcc and bcc is 0.036eV for Cu by Domain
and Becquart\cite{PhysRevB.65.024103} and 0.10eV for Ni by Mishin
\textit{et al.}\cite{Mishin20054029}, though these small differences
are ignored in fitting.

\begin{table}
\caption{Comparison of target \textit{ab-intio} thermodynamic properties and
fitted results in RLM\label{tab:Comparison-of-energies} . Target
values are from Ref. \onlinecite{Vincent2008} (except the marked
ones)}
\begin{ruledtabular}

\begin{tabular}{cccccc}
 &
target &
fitted &
 &
target &
fitted\tabularnewline
\hline 
$E_{Cu}^{sol}\left(\mathrm{Fe}\right)$ &
\ 0.55 &
\ 0.5562 &
$E_{CuNi}^{b\left(1\right)}\left(\mathrm{Fe}\right)$ \footnotemark[2] &
\ 0.065 &
\ 0.0661\tabularnewline
$E_{Ni}^{sol}\left(\mathrm{Fe}\right)$ &
-0.17 &
-0.1804 &
$E_{CuNi}^{b\left(2\right)}\left(\mathrm{Fe}\right)$ \footnotemark[2] &
0.02 &
\ 0.0258\tabularnewline
$E_{CuCu}^{b\left(1\right)}\left(\mathrm{Fe}\right)$ &
\ 0.16 &
\ 0.1075 &
$E_{NiNi}^{b\left(1\right)}\left(\mathrm{Fe}\right)$ &
-0.10\  &
-0.0440\tabularnewline
$E_{CuCu}^{b\left(2\right)}\left(\mathrm{Fe}\right)$ &
\ 0.05 &
\ 0.0421 &
$E_{NiNi}^{b\left(2\right)}\left(\mathrm{Fe}\right)$ &
-0.02\  &
-0.0015\tabularnewline
$E_{\mathrm{V}}^{for}\left(\mathrm{Fe}\right)$ &
\ 2.00 &
\ 1.9658 &
$E_{Cu\mathrm{V}}^{b\left(1\right)}\left(\mathrm{Fe}\right)$ \footnotemark[3] &
0.16 &
\ 0.0897\tabularnewline
$E_{\mathrm{V}}^{for}\left(\mathrm{Cu}\right)$ \footnotemark[1] &
\ 1.05 &
\ 1.0412 &
$E_{Cu\mathrm{V}}^{b\left(2\right)}\left(\mathrm{Fe}\right)$ \footnotemark[3] &
0.18 &
\ 0.1272\tabularnewline
$E_{\mathrm{V}}^{for}\left(\mathrm{Ni}\right)$ &
\ 0.60 &
\ 0.5993 &
$E_{Ni\mathrm{V}}^{b\left(1\right)}\left(\mathrm{Fe}\right)$ &
0.03 &
\ 0.0241\tabularnewline
 &
 &
 &
$E_{Ni\mathrm{V}}^{b\left(2\right)}\left(\mathrm{Fe}\right)$ &
0.17 &
\ 0.1656\tabularnewline
\end{tabular}

\end{ruledtabular}

\footnotetext[1] {The original value is 1.6eV, significantly higher than ab-intio result by Soisson \textit{et al.} \cite{PhysRevB.76.214102}, also higher than prediction by interatomic potentials\cite{doi:10.1080/01418619708207198,Pasianot2007118,0965-0393-20-4-045016}. So we modified it.}

\footnotetext[2] {Fitting of original value gives incorrect number density, i.e. density decrease with increasing Ni content, so modified intentionally to a level according to the ternary potential of Bonny \textit{et al.}\cite{doi:10.1080/14786430903299337}.}

\footnotetext[3] {Modified according to Ref. \onlinecite{PhysRevB.65.024103}.}
\end{table}

As suggested by Soisson and Fu\cite{PhysRevB.76.214102}, it seems
necessary to introduce non-configurational entropy into the energy
model to get better solubility of Cu, which is done by adding temperature
dependent term for the Fe-Cu bond energies. We also used this method.
Since saddle state is more related to initial state, this modification
is only introduced into initial system energies. So $E_{ini}$ becomes
$E_{ini}^{\prime}$ with its pairwise bonds modified to $\varepsilon_{X-Y}^{\left(i\right)\prime}=\varepsilon_{X-Y}^{\left(i\right)}-\lambda_{X-Y}^{\left(i\right)}T$.
 The \textit{\textgreek{l}} values for Fe-Cu pair are obtained by
fitting Cu solubility in $\alpha$\textendash{}Fe, corresponding non-configurational
entropy $\triangle S_{nc}$ equals 1.3\textit{k}\textsubscript{B},
for Fe-Ni and \textit{X}-V pairs, $\triangle S_{nc}$ were set as
1.0\textit{ k}\textsubscript{B}. The \textit{\textgreek{l}} factors
for related pairs are listed in Table \ref{tab:Entropy-contribution}.\nocite{Miloudi1997,salje:1833}

\begin{table}

\caption{Entropy contribution factor \textit{$\lambda$} for pairwise bonds\label{tab:Entropy-contribution}}

\begin{ruledtabular}

\begin{tabular}{ccc}
\multirow{2}{*}{\textcolor{black}{Bond }\textit{\textcolor{black}{X}}\textcolor{black}{-}\textit{\textcolor{black}{Y}}} &
\multicolumn{2}{c}{$\lambda_{X-Y}^{\left(i\right)}$}\tabularnewline
\cline{2-3} 
 & \textit{i}=1 &
\textit{i}=2\tabularnewline
\hline 
Fe-Cu &
8.8450\texttimes{}10\textsuperscript{-6} &
6.8774\texttimes{}10\textsuperscript{-6}\tabularnewline
Fe-Ni &
6.9518\texttimes{}10\textsuperscript{-6} &
5.0930\texttimes{}10\textsuperscript{-6}\tabularnewline
Fe-V &
7.8816\texttimes{}10\textsuperscript{-6} &
3.8534\texttimes{}10\textsuperscript{-6}\tabularnewline
Cu-V &
6.5202\texttimes{}10\textsuperscript{-6} &
5.6686\texttimes{}10\textsuperscript{-6}\tabularnewline
Ni-V &
5.9914\texttimes{}10\textsuperscript{-6} &
6.3738\texttimes{}10\textsuperscript{-6}\tabularnewline
\end{tabular}

\end{ruledtabular}
\end{table}

\subsection{Time adjusting}

The time adjusting has special importance for AKMC thermal ageing
simulations applied to solid solutions. For one reason the computation
cost requires relatively smaller simulation box which usually causing
significant higher vacancy concentration and under-estimated time
evolution. Moreover, phase separation progress results in evolving
equilibrium vacancy concentration, while number of vacancies introduced
in AKMC is fixed and integral. This difference cause time under-estimate
degree evolves nonlinearly. We used the model described in Ref. \onlinecite{PhysRevB.65.094103},
physical time is rescaled by the ratio between vacancy concentration
within matrix in MC model and equilibrium vacancy concentration of
matrix,
\begin{equation}
\mathrm{d}t_{\mathrm{real}}=\mathrm{d}t_{\mathrm{MC}}\cdot\frac{C_{\mathrm{V}}^{\mathrm{MC}}\left(\mathrm{M}\right)}{C_{\mathbf{V}}^{\mathrm{eq}}\left(\mathrm{M}\right)},\label{eq:integration_of_time}
\end{equation}
Equilibrium vacancy concentration of matrix is determined by vacancy
formation energy in matrix, $C_{\mathrm{V}}^{\mathrm{eq}}\left(\mathrm{M}\right)=\exp\left(\frac{\triangle S}{k_{\mathrm{B}}}\right)\exp\left[-\frac{E_{\mathrm{V}}^{for}\left(\mathrm{M}\right)}{k_{\mathrm{B}}T}\right]$,
$\triangle S$ is taken as 1.0\textit{ k}\textsubscript{B} as we
mentioned above. Vacancy concentration
within matrix in MC model is given by,
\begin{equation}
C_{\mathrm{V}}^{\mathrm{MC}}\left(\mathrm{M}\right)=\frac{f_{\mathrm{V}}^{\mathrm{M}}}{NX_{\mathrm{M}}},
\end{equation}
where, \textit{X}\textsubscript{M} is the concentration of matrix
atoms(Fe) in the box, \textit{N} is the total atom number. $f_{\mathrm{V}}^{\mathrm{M}}$
is the ratio between equilibrium vacancy concentration in the solid
solution and concentration in pure matrix.

Our concern focuses on the $f_{\mathrm{V}}^{\mathrm{M}}$ value. There
has not been a physical analysis for this value. In Ref. \onlinecite{PhysRevB.76.214102,PhysRevB.65.094103},
it is regarded as fraction of time spent by the vacancy in Fe matrix,
thus in their simulations the value was computed by checking if first
and second nearest neighbors of vacancy contain Cu atom. One difficulty
of this method is that because the system evolves nonlinearly with
time, the ideal step length is nearly impossible to be obtained \textit{on
the fly}. As a consequence, the slope of $f_{\mathrm{V}}^{\mathrm{M}}\left(t\right)$
curve is slightly deviated at the beginning of simulation, and data
fluctuation continuing become remarkable stronger with MC time increasing,
especially in the coarsening stage. Here, we propose a semi-empirical
way to get the $f_{\mathrm{V}}^{\mathrm{M}}$ value. According to
Lomer\cite{Lomer1958}, for a dilute solid solution, the value for
ideal solution state can be determined physically, as well as the
state when precipitation completed. Soisson and Fu\cite{PhysRevB.76.214102}
 have given the expressions for Fe\textendash{}Cu
system,
\begin{equation}
f_{\mathrm{V}}^{\mathrm{M}}\left(0\right)=\frac{1-z_{1}C_{\mathrm{Cu}}-z_{2}C_{\mathrm{Cu}}}{\left(1-z_{1}C_{\mathrm{Cu}}-z_{2}C_{\mathrm{Cu}}\right)+\underset{i=1}{\overset{2}{\sum}}z_{i}C_{\mathrm{Cu}}\exp\left(\frac{E_{\mathrm{CuV}}^{b\left(i\right)}\left(\mathrm{Fe}\right)}{k_{\mathrm{B}}T}\right)},
\end{equation}
\begin{equation}
f_{\mathrm{V}}^{\mathrm{M}}\left(\infty\right)=1/\left[1+\frac{C_{\mathrm{Cu}}}{1-C_{\mathrm{Cu}}}\exp\left(\frac{E_{\mathrm{V}}^{for}\left(\mathrm{Fe}\right)-E_{\mathrm{V}}^{for}\left(\mathrm{Cu}\right)}{k_{\mathrm{B}}T}\right)\right],
\end{equation}

The expressions give the upper and lower bounds for the $f_{\mathrm{V}}^{\mathrm{M}}$
value. Since $f_{\mathrm{V}}^{\mathrm{M}}$ decrease monotonically,
an S-shape sigmoidal function is expected to link the bounds. The
cluster number density, mean cluster size and advancement factor,
also evolve with time, there should be a mapping relationship between
the evolution of $f_{\mathrm{V}}^{\mathrm{M}}$ and those quantities
on the basis of time. Since the mentioned physical quantities can
be calculated in post-processing, if a sigmoidal function and mapping
relationship of several special points is given, a semi-empirical
function of $f_{\mathrm{V}}^{\mathrm{M}}$ evolution can be obtained.

Since it is known evolution of advancement factor follows the Johnson\textendash{}Mehl\textendash{}Avrami
(JMA) law, $\xi\left(t\right)=1-\exp\left[-\left(t/\tau\right)^{n}\right]$.
$f_{\mathrm{V}}^{\mathrm{M}}$ should evolve no slower than the order
of JMA law. The sigmoidal Hill equation\cite{Hill1910} has been chosen
to represent the $f_{\mathrm{V}}^{\mathrm{M}}$ evolution, the formula
is given as below,
\begin{equation}
\frac{\log\left(f\right)-\log\left(f_{0}\right)}{\log\left(f_{\infty}\right)-\log\left(f_{0}\right)}=\frac{1}{1+\left(t_{\mathrm{real}}/t_{0}\right)^{p}},
\end{equation}
where \textit{t}\textsubscript{real} is the real time, \textit{t}\textsubscript{0}
and \textit{p} are unknown parameters.

Use a substitution $g=\log\left(t_{\mathrm{real}}\right)$, we get
the following,
\begin{equation}
\frac{\log\left(f\right)-\log\left(f_{0}\right)}{\log\left(f_{\infty}\right)-\log\left(f_{0}\right)}=\frac{1}{1+\exp\left[\left(g-g_{0}\right)/dx\right]},\label{eq:boltz}
\end{equation}
where $g_{0}=\log\left(t_{0}\right)$ and $dx=1/p$.

The expression on right of eq. \prettyref{eq:boltz} is known as Boltzmann
equation. The $f_{\mathrm{V}}^{\mathrm{M}}\left(t\right)$ calculated
in Ref. \onlinecite{PhysRevB.76.214102} got perfect fitting by this
equation, with $g_{0}\cong-0.82$ and $dx\cong0.25$ (using 10 as
the base of log function). This implies the eq. \prettyref{eq:boltz}
has the correct \textquotedblleft{}evolving speed\textquotedblright{}.
So the next step to get the semi-empirical solution for $f_{\mathrm{V}}^{\mathrm{M}}$
is to fit the unknown parameters of \textit{g}\textsubscript{0} and
\textit{dx}.

\begin{figure}
\includegraphics[width=7cm]{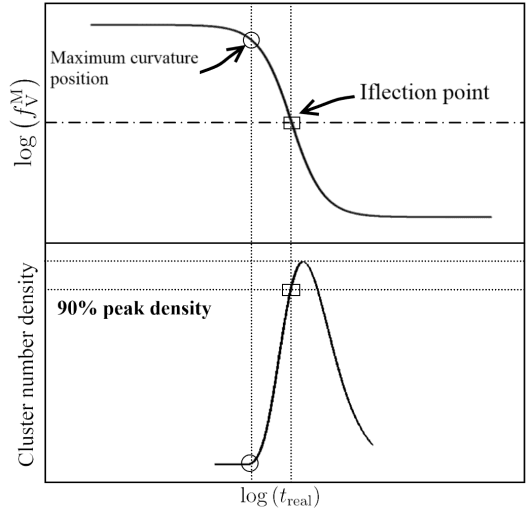}

\caption{Real time fitting scheme by relationship between number density and
$f_{\mathrm{V}}^{\mathrm{M}}$.\label{fig:Real-time-fitting}}
\end{figure}

We have chosen the number density of Cu clusters to give a mapping relationship of
special points between post-processing data and $f_{\mathrm{V}}^{\mathrm{M}}$.
The reason is that this variable does not evolve monotonically which
making distinguish of different stages easily. On the number density
evolution, two regions have the best analysis properties making them
special. One is the time when nucleation starts, the other is the
growth stage. When nucleation starts, binding of vacancy to clusters
should has the most significant change, since random embryos from
thermal fluctuation become stable nucleuses, this will cause a rapid
change of the slope of the $f_{\mathrm{V}}^{\mathrm{M}}\left(t\right)$
curve. So we assumed the time nucleation starts , identified by when
number density starts increasing, is mapped to the maximum curvature
position ($\frac{\mathrm{d^{4}}f_{\mathrm{V}}^{\mathrm{M}}}{\mathrm{d}t^{4}}=0$)
of $f_{\mathrm{V}}^{\mathrm{M}}\left(t\right)$ curve, that is $\left(g_{0}+dx\cdot\ln\left(5-2\sqrt{6}\right),\left(3+\sqrt{6}\right)/6\right)$
for Boltzmann equation. Similarly, at growth stage, new nucleus stops
to form, a maximum slope is expected since contribution to vacancy
binding from nucleation is lost. We assumed the time growth starts
mapped to the inflection point ($\frac{\mathrm{d^{2}}f_{\mathrm{V}}^{\mathrm{M}}}{\mathrm{d}t^{2}}=0$),
that is $\left(g_{0},0.5\right)$, and identified by number density
first reaches 90\% of peak density. The mapping relationship is represented
in Fig. \ref{fig:Real-time-fitting}. MC time of those two positions
can be directly read from number density evolution, namely $t_{\mathrm{MC}}^{n}$
for time when nucleation starts and $t_{\mathrm{MC}}^{g}$ for time
when growth starts. According to eq. \prettyref{eq:integration_of_time},
the MC time can be obtained by integration as a function of real time.
Thus by solving the problem of equations \prettyref{eq:tmc_n} and
\prettyref{eq:tmc_g} with proper optimization method, \textit{g}\textsubscript{0}
and \textit{dx} can be fitted. Corresponding relationship between
each MC time and real time can be obtained by numerical integration
within an acceptable margin of error. This semi-empirical method is
based on the post-processing of the number density data from AKMC
and avoids complicate record step length setting and data fluctuation.
The obtained $f_{\mathrm{V}}^{\mathrm{M}}\left(t\right)$ relationship
is nonlinear as the vacancy bind energy evolution.

\begin{equation}
NX_{\mathrm{M}}C_{\mathbf{V}}^{\mathrm{eq}}\left(\mathrm{M}\right)\int_{0}^{nuleation-start}\left(1/f_{\mathrm{V}}^{\mathrm{M}}\right)\mathrm{d}t_{\mathrm{real}}=t_{\mathrm{MC}}^{n},\label{eq:tmc_n}
\end{equation}
\begin{equation}
NX_{\mathrm{M}}C_{\mathbf{V}}^{\mathrm{eq}}\left(\mathrm{M}\right)\int_{nuleation-start}^{growth-start}\left(1/f_{\mathrm{V}}^{\mathrm{M}}\right)\mathrm{d}t_{\mathrm{real}}=t_{\mathrm{MC}}^{g}-t_{\mathrm{MC}}^{n}\label{eq:tmc_g}
\end{equation}

\section{Results}

\subsection{Cluster identification}

Cu rich clusters are identified by counting number of \textquotedblleft{}bonds\textquotedblright{}
linked to lattice sites. Those bonds are ranged up to second nearest
neighbors as the activation energy model defined. Since Fe atom is
absent inside clusters in our results, which consist with other simulations,
Fe atoms are ignored during the identification. Because of the time
adjusting method we used, a lesser identify rule has been used, to
be recognized as part of a cluster, one atom is required to be linked
with at least three bonds. This geometrically causes the smallest
cluster recognized is a tetrahedron.

\subsection{Fe\textendash{}Cu binary system kinetics}

We applied the model above to simulate the precipitation kinetics
of a Fe\textendash{}1.34 at. \% Cu alloy (about 1.4 wt. \% Cu), during
thermal ageing at four different temperatures, 663K, 713K, 773K
and 873K respectively.

\begin{figure}

\includegraphics[width=8cm]{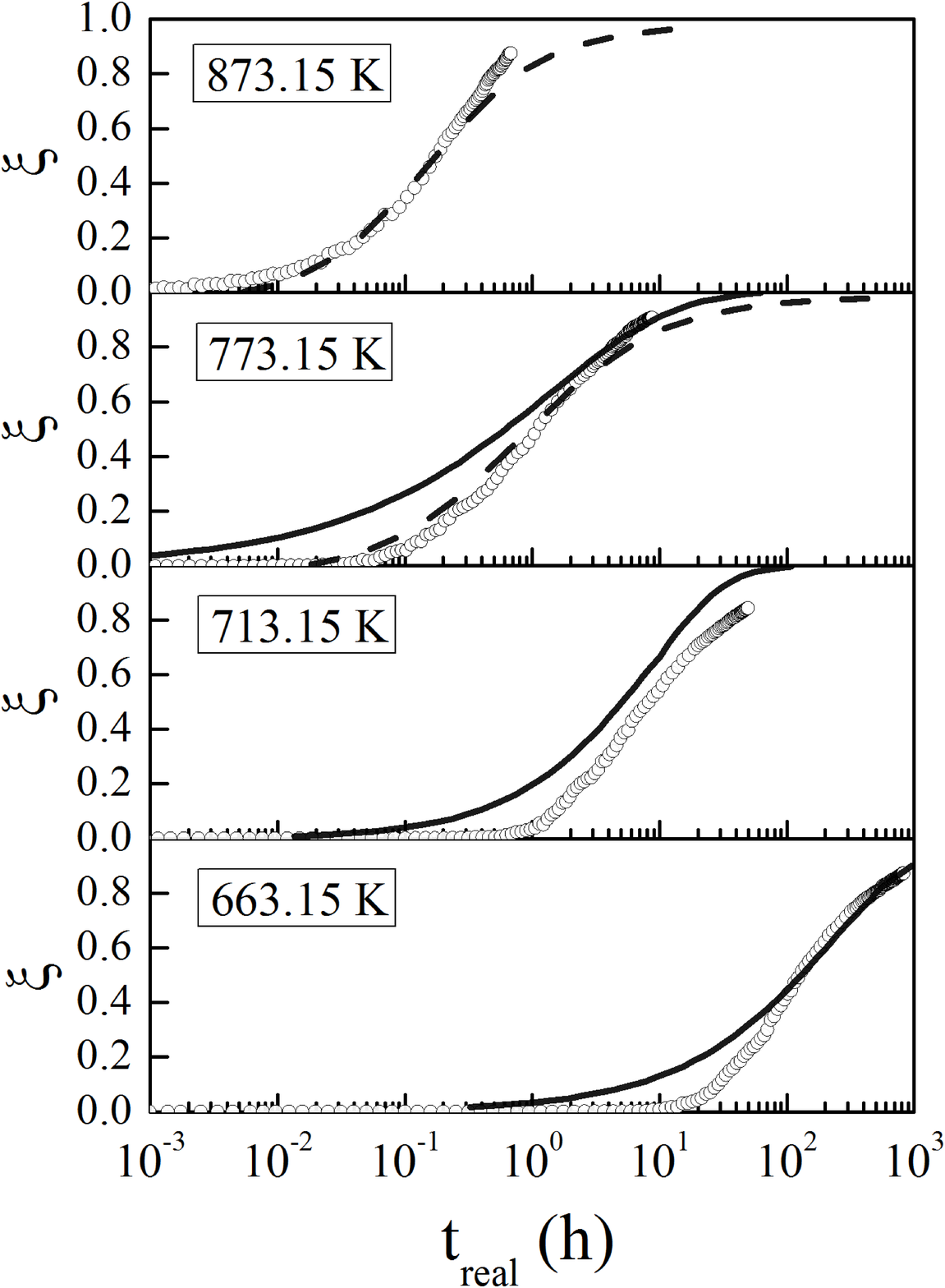}\caption{Evolution of the advancement factor of Fe\textendash{}1.34 at. \%
Cu under different temperatures.\label{fig:advancement_factor}}
\end{figure}

The advancement factor \textit{$\xi$}, defined by eq. \prettyref{eq:advancement_factor}
, represents the completeness of precipitation progress, and is a
basic property which can be measured by several characteristics techniques.
\begin{equation}
\xi\left(t\right)=\frac{C_{\mathrm{Cu}}\left(0\right)-C_{\mathrm{Cu}}\left(t\right)}{C_{\mathrm{Cu}}\left(0\right)-C_{\mathrm{Cu}}\left(\infty\right)},\label{eq:advancement_factor}
\end{equation}
in the equation, \textit{C}\textsubscript{Cu} is the concentration
of solute Cu atoms. \textit{C}\textsubscript{Cu}(0) is the initial
Cu content, and \textit{C}\textsubscript{Cu}($\infty$) is the solubility
of Cu in $\alpha$\textendash{}Fe which is estimated by the following
equation,
\begin{equation}
C_{\mathrm{Cu}}\left(\infty\right)\doteq\exp\left(\frac{\triangle S_{nc}^{Cu}}{k_{\mathrm{B}}}\right)\exp\left[\frac{E_{\mathrm{Cu}}^{sol}\left(\mathrm{Fe}\right)}{k_{\mathrm{B}}T}\right],
\end{equation}
where, $\triangle S_{nc}^{Cu}$ is the non-configurational entropy,
and $E_{\mathrm{Cu}}^{sol}\left(\mathrm{Fe}\right)$ is the mixing
energy of Cu. 
To obtain the advancement factor, a simulation box containing 64\texttimes{}64\texttimes{}64 bcc
unit cells was used. The evolution of the advancement factor is shown
in Fig. \ref{fig:advancement_factor}. The curves have good agreement
with experiments overall the temperature range. Exceptions are, at
low temperatures 663K and 713K, the kinetics are slower than
experiment at the beginning of precipitation, and at high temperatures
773K and 873K, the kinetics are faster than experiments when
\textit{$\xi>0.6$. }Even though, several simulation tests in bigger
boxes (128\texttimes{}128\texttimes{}128) with \textit{$\xi$ }up
to 0.7 show large box size seems to have better agreement. And, using
more strict identification rule, i.e. cluster no smaller than 10 atoms,
the kinetics will only be slightly slower at the beginning.

\begin{figure}
\includegraphics[width=8cm]{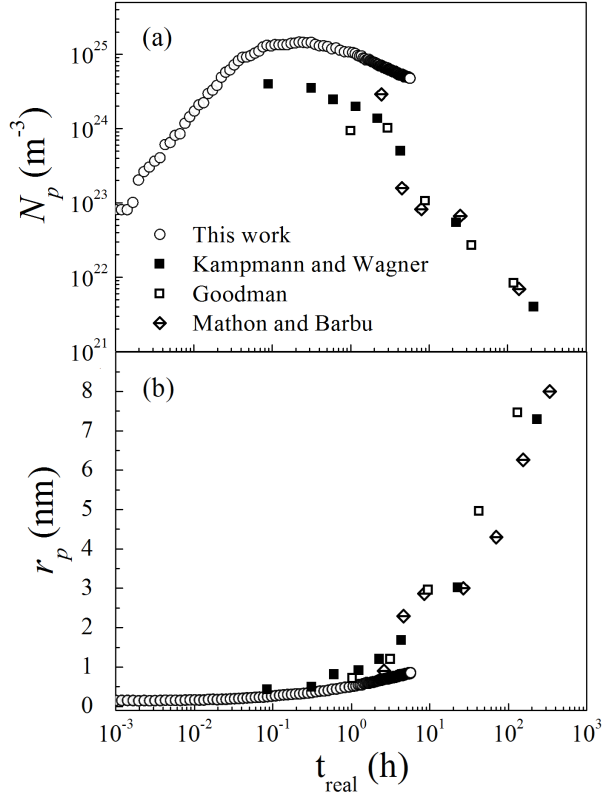}\caption{Precipitation kinetics of Fe\textendash{}1.34 at. \% Cu under 773K,
(a) cluster number density evolution $N_{p}\left(t\right)$, (b) mean
cluster radius evolution $r_{p}\left(t\right)$. Experimental results
are from Kampmann and Wagner: Ref. \onlinecite{Kampmann1986}, Goodman:
Ref. \onlinecite{Goodman1973a}, Mathon and Barbu: Ref. \onlinecite{Mathon1997224} .\label{fig:number_density_binary}}
\end{figure}

Advancement factor is still not enough to reflect all aspects of precipitation
kinetics. A simulation box of 128\texttimes{}128\texttimes{}128 unit
cells has been used to get the number density and the mean cluster
radius of Cu clusters. As is shown in Fig. \ref{fig:number_density_binary}(a),
the number density and mean cluster size evolution has similar tendency
of experiments. But the absolute value of the calculated number density
is nearly a magnitude of order higher than the experiment data. In
our AKMC result, after 10\textsuperscript{0} h clusters are expected
to have larger size than experiment detection limit, since the coarsening
has already started. However, the number density after 10\textsuperscript{0}
h is still higher than experiments. So this kind of deviate is more
than identification difference. To our knowledge, Castin \textit{et
al.}\cite{castin:064502} gave the most satisfied result by hybrid
AKMC method with ANN trained by an embedded atomic (EAM) potential\cite{Pasianot2007118}.
According to classical nucleation theory, the critical nucleation
size is determined by volume free-energy change and interfacial energy
of Cu cluster. The interfacial energies evaluated by our pairwise
bonds are $E_{\mathrm{FeCu}}^{\mathrm{int}\left\{ 100\right\} }$=0.5035J/m\textsuperscript{2},
$E_{\mathrm{FeCu}}^{\mathrm{int}\left\{ 110\right\} }$=0.4142J/m\textsuperscript{2}
and $E_{\mathrm{FeCu}}^{\mathrm{int}\left\{ 111\right\} }$=0.4465
J/m\textsuperscript{2}. And molecular statics calculation shows the
EAM potential from Ref. \onlinecite{Pasianot2007118} gives $E_{\mathrm{FeCu}}^{\mathrm{int}\left\{ 100\right\} }$=0.3789
J/m\textsuperscript{2} $E_{\mathrm{FeCu}}^{\mathrm{int}\left\{ 110\right\} }$=0.4113J/m\textsuperscript{2}
and $E_{\mathrm{FeCu}}^{\mathrm{int}\left\{ 111\right\} }$=0.5130J/m\textsuperscript{2}.
The interfacial energies evaluated by both methods are close. However,
as we mentioned before, the binding energies of Cu atoms are significantly
under-estimated by pairwise bonds, the volume free-energy change should
also be under-estimated. Thus the critical nucleation size will be
under-estimated, result in a higher peak number density, as well as
the density in coarsening stage.

\begin{figure}
\includegraphics[width=8cm]{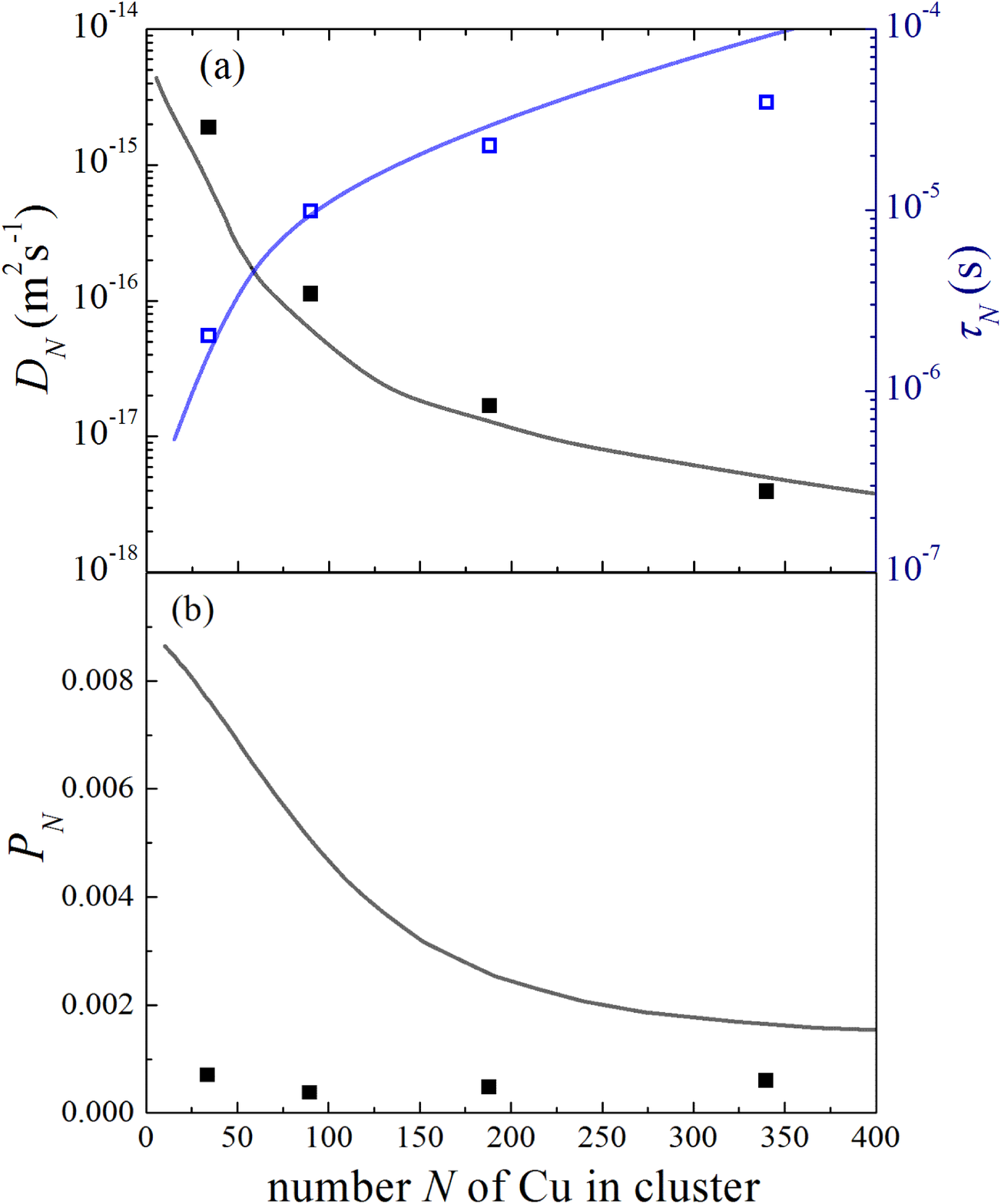}\caption{Cluster mobility of VCu\textit{\textsubscript{\textit{N}}} under
773K, (a) diffusion coefficients $D_{N}$ and lifetime $\tau_{N}$
versus cluster number (b) dissolution probability $P_{N}$ versus
cluster number. Squares: this work; line: \textit{B}-spline line of
data from Ref.\onlinecite{castin:064502}.\label{fig:Cluster-mobility-binary}}
\end{figure}

It was first suggested by Soisson \textit{et al.}\cite{PhysRevB.76.214102,Soisson19963789}
that the Cu precipitation in $\alpha$\textendash{}Fe may favor a coagulation
mechanism caused by highly mobile Cu clusters. This theory was recently
confirmed by a hybrid AKMC method\cite{castin:064502}. We noticed
that the difference between coagulation and emitting-absorbing mechanism
only become marked in long term coarsening stage, our AKMC results
seems to be not strong enough to be proved \textquotedblleft{}correct\textquotedblright{}.
It was mentioned that their Object KMC part is controlled by several parameters of Cu clusters,
i.e. lifetime, diffusion coefficients and dissolution probability.
Therefore we used the AKMC simulated mobility of VCu\textit{\textsubscript{\textit{N}}}
clusters at 773K. Fig. \ref{fig:Cluster-mobility-binary} shows
the diffusion coefficients, lifetime and dissolution probability versus
time, each data point was simulated for tens of thousands of times
for statistics. The diffusion coefficients and lifetime have a good
agreement with Ref. \onlinecite{castin:064502}. The dissolution probability
by our model shows a different pattern, it does not decrease quickly
with cluster size. However, the dissolution probability calculated by us is always
lower than the probability in Ref. \onlinecite{castin:064502}. It is
apparent only a system with large dissolution probability will tend
to emitting-absorbing mechanism. According to their work, even if
dissolution is forbidden which means a dissolution probability of
0, the evolution still has good agreement with slight larger cluster
size compared to experiments. So our model actually can reproduce
the coagulation mechanism, rather than emitting-absorbing mechanism.

It has been shown from above results, our model reproduces
 consistent results with experiments and other simulations over a temperature range. The advancement factor
was reproduced well, as well as the Cu cluster mobility. Although
the cluster density is over-estimated by our model, it is a drawback
of pairwise bonds. Nevertheless, good description of Cu precipitation
kinetics in binary Fe\textendash{}Cu system has been obtained by our
AKMC model and combined time adjusting method, the very first step
needed for the study of Ni effect.

\subsection{Fe\textendash{}Cu\textendash{}Ni ternary system kinetics}

We performed thermal ageing simulations at 823K on Fe\textendash{}1.34 at. \% Cu
\textendash{}\textit{x} at. \% Ni alloys, the \textit{x} values are
0, 0.5, 1.0, 1.5 and 2.0, respectively.
The simulations were conducted in a box containing 128\texttimes{}128\texttimes{}128
unit cells.

By direct visual observation from snapshot, we found that Ni appears
to remain random distribution during the simulation, though some Ni
atoms embellishes on Cu clusters like strawberry. Based on this fact,
in the ternary alloys, one more rule was added to cluster identification,
that is clusters should have only one pure Cu core no smaller than
a tetrahedron. And, the bounds of $f_{\mathrm{V}}^{\mathrm{M}}$ is
extended as follows,
\begin{equation}
f_{\mathrm{V}}^{\mathrm{M}}\left(0\right)=\frac{\underset{X=Cu,Ni}{\sum}\left(1-z_{1}C_{\mathrm{X}}-z_{2}C_{\mathrm{X}}\right)}{\underset{X=Cu,Ni}{\sum}\left[\left(1-z_{1}C_{\mathrm{X}}-z_{2}C_{\mathrm{X}}\right)+\overset{2}{\underset{i=1}{\sum}}z_{i}C_{\mathrm{X}}\exp\left(\frac{E_{\mathrm{XV}}^{b\left(i\right)}\left(\mathrm{Fe}\right)}{k_{\mathrm{B}}T}\right)\right]},
\end{equation}
\begin{equation}
f_{\mathrm{V}}^{\mathrm{M}}\left(\infty\right)=\frac{1-z_{1}C_{\mathrm{Ni}}-z_{2}C_{\mathrm{Ni}}-C_{\mathrm{Cu}}}{\left(1-C_{\mathrm{Cu}}-\overset{2}{\underset{i=1}{\sum}}z_{i}C_{\mathrm{Ni}}\right)+C_{\mathrm{Cu}}\exp\left(\frac{E_{\mathrm{V}}^{for}\left(\mathrm{Fe}\right)-E_{\mathrm{V}}^{for}\left(\mathrm{Cu}\right)}{k_{\mathrm{B}}T}\right)+\overset{2}{\underset{i=1}{\sum}}z_{i}C_{\mathrm{Ni}}\exp\left(\frac{E_{\mathrm{NiV}}^{b\left(i\right)}\left(\mathrm{Fe}\right)}{k_{\mathrm{B}}T}\right)},
\end{equation}

\begin{figure}
\includegraphics[width=8cm]{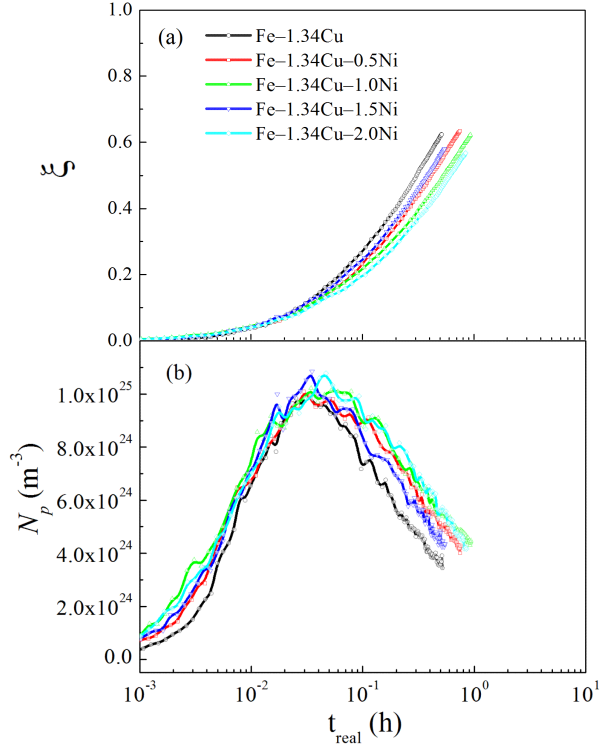}\caption{Precipitation of Fe\textendash{}Cu\textendash{}Ni ternary alloys under
823K, (a) advancement factor evolution $\xi\left(t\right)$, (b)
cluster number density evolution $N_{p}\left(t\right)$.\label{fig:Precipitation-of-Fe=002013Cu=002013Ni}}
\end{figure}

The advancement factor evolution is shown in Fig. \ref{fig:Precipitation-of-Fe=002013Cu=002013Ni}(a).
Starting from about 2\texttimes{}10\textsuperscript{-2}h, the precipitation
kinetics become slower in four ternary alloys. When refer to Fig.
\ref{fig:Precipitation-of-Fe=002013Cu=002013Ni}(b) of number density
evolution, we can see that, before number density reaches the peak
number density the evolution of number density in ternary alloys is
even slight faster. However, when passed the peak number density position,
the evolution of number density in ternary alloys become obviously
slower than the binary Fe\textendash{}Cu alloy. The time 2\texttimes{}10\textsuperscript{-2}h
actually corresponds to the time at peak density. So from these observations
it has been confirmed that the precipitation kinetics of ternary alloys
appears to be \textquotedblleft{}delayed\textquotedblright{} in the
coarsening stage.

At the peak density point, the ternary alloys containing 1.5 at. \% and 2.0
at. \% Ni have about 10\% higher number density than the binary alloy,
while the other two ternary alloys are about 3\% higher. It was found
the peak number density of Cu clusters in Fe\textendash{}1.13 at. \% Cu\textendash{}1.36 at. \% Ni
is about 29\% higher than Fe\textendash{}1.13 at. \% Cu by ANN AKMC in Ref.
\onlinecite{doi:10.1080/14786430903299824}. And the experimental
results by Buswell \textit{et al.}\cite{Buswell-1990} gave about
34\% higher peak number density by means of larger area under density-size
distribution. Our result reproduced a similar tendency. The increasing
peak number density is possibly caused by Ni effect on nucleation.
Al-Motasem \textit{et al.}\cite{AlMotasem2011215} revealed Cu\textit{\textsubscript{\textit{m}}}Ni\textit{\textsubscript{\textit{n}}}
clusters have lower formation energy than pure Cu clusters. Seko \textit{et
al.}\cite{SEKO2004} also gave a similar result by \textit{ab-initio}
calculation. So the reason for a higher peak number density of Cu
clusters is that Ni atoms act as nucleation centers for Cu precipitates
and promote nucleation of Cu clusters. This also explains the more
rapid evolution of ternary alloys during nucleation stage.

\begin{figure}

\includegraphics[width=8cm]{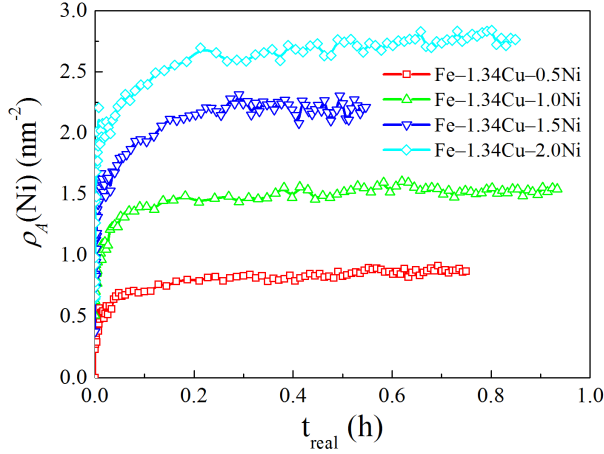}\caption{The evolution of surface area density of Ni atoms on clusters $\rho_{A}\left(\mathrm{Ni}\right)\left[t\right]$.\label{fig:rho_Ni}}

\end{figure}

The mean surface area density of Ni atoms surrounding Cu cluster is shown
in Fig. \ref{fig:rho_Ni}. The surface area density of Ni increases
quickly with time and then saturates. The more Ni contains in a ternary
alloy, the higher saturated surface area density of Ni is formed.
Actually, a linear relationship has been found between the saturated
surface area density of Ni and the Ni content of the alloy, with a
slope around 1.46 nm\textsuperscript{-2}/(1\%Ni). The figure also
confirmed visual observation that few Ni atoms embellish on Cu clusters,
since the density is quite low.

\section{Discussion}

\begin{figure}

\begin{minipage}[t]{1\columnwidth}%
\includegraphics[width=8cm]{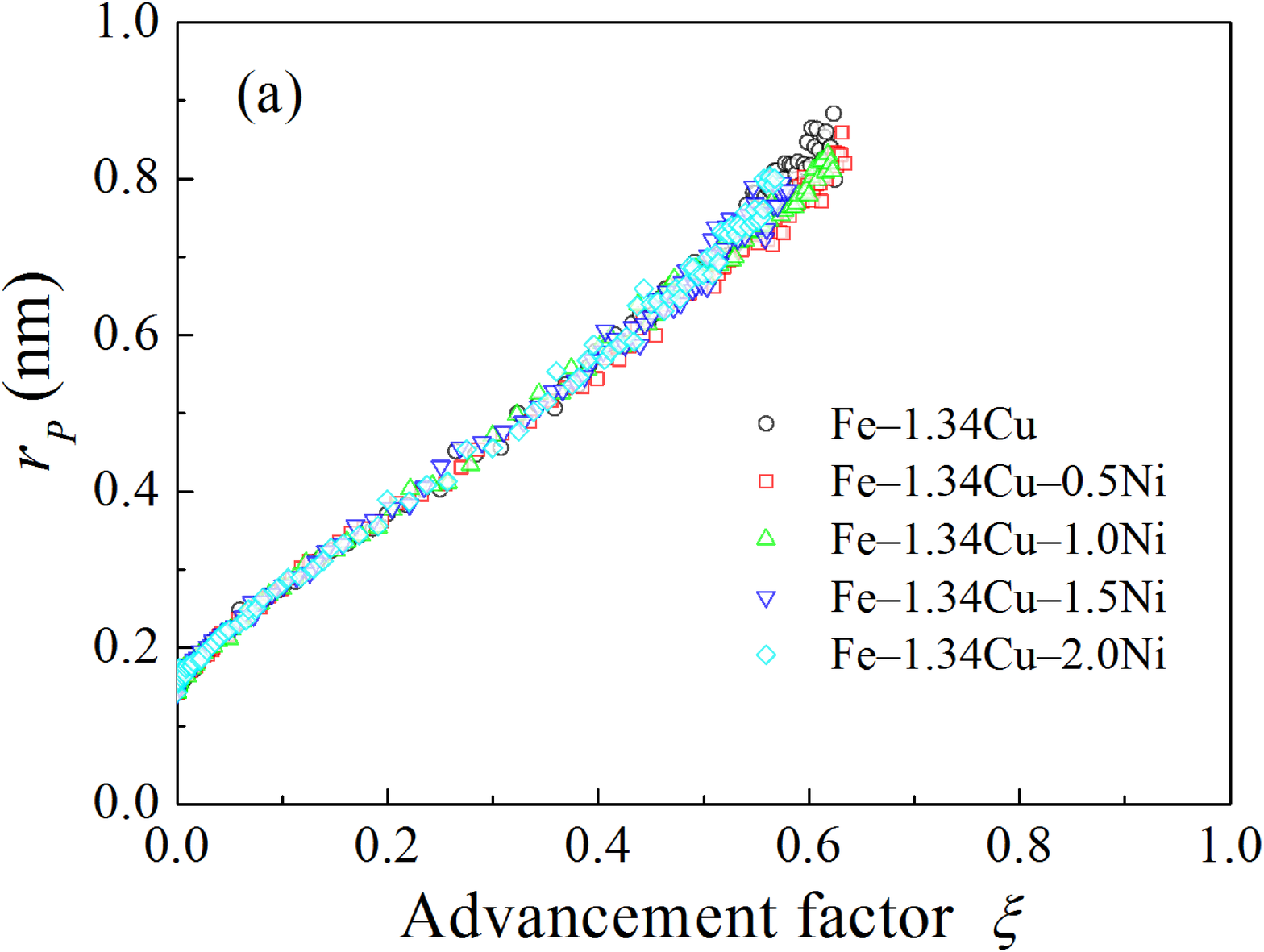}%
\end{minipage}\vfill{}
\begin{minipage}[t]{1\columnwidth}%
\includegraphics[width=8cm]{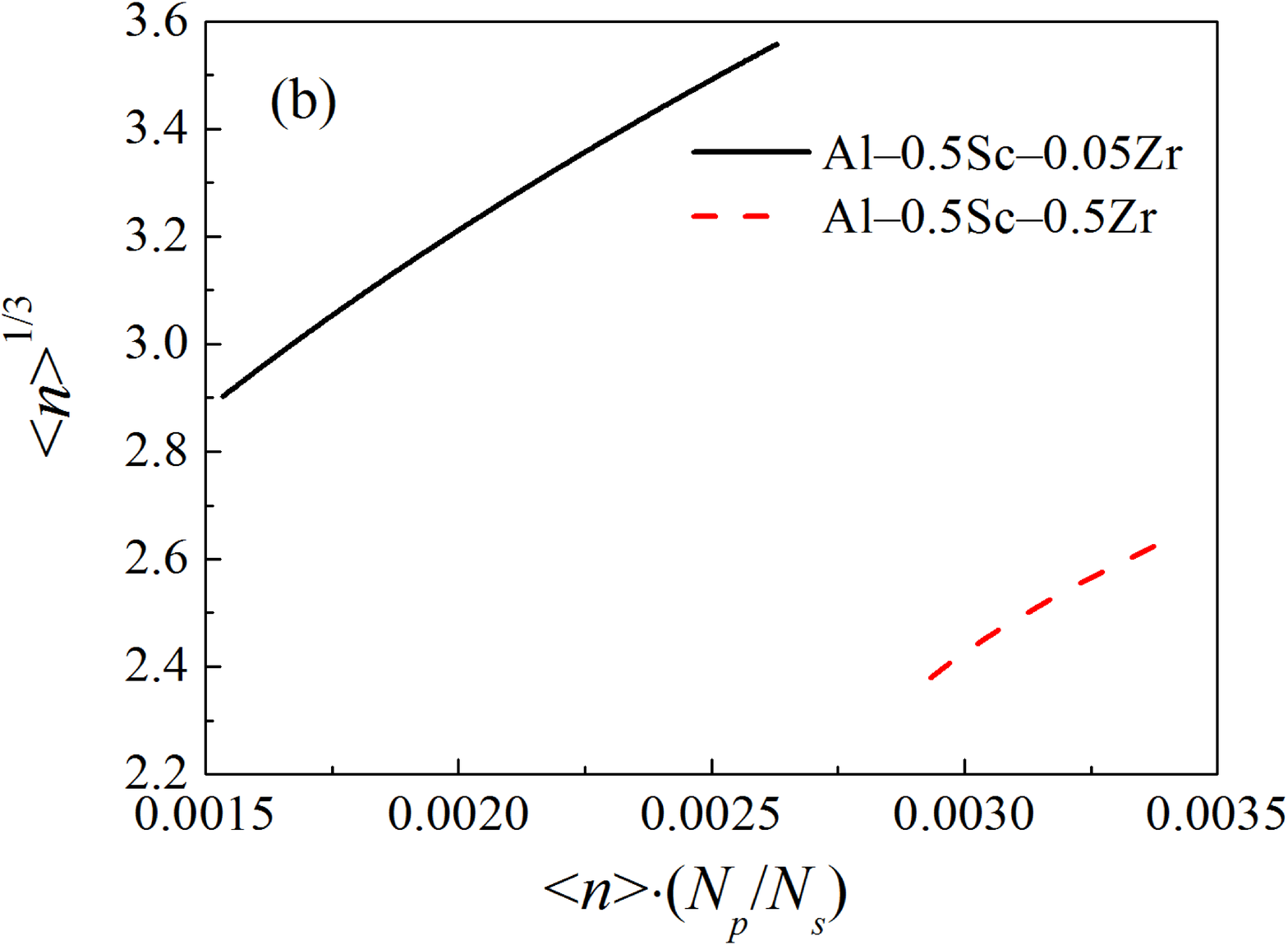}%
\end{minipage}\caption{Comparison of addition effect of third-party element on precipitation
kinetics in view of mean size versus cluster density, (a) Ni effect
in Fe\textendash{}Cu\textendash{}Ni alloys, (b) Zr effect in Al\textendash{}Sc\textendash{}Zr
alloys.\label{fig:Comparison}}
\end{figure}

One obvious effect of Ni on Cu precipitation is that higher cluster number density found in ternary alloys
during coarsening stage compared to binary Fe\textendash{}Cu alloy. In other
alloy systems, similar phenomenon caused by third-party element atoms
can also be found. For example, in Al\textendash{}Sc\textendash{}Zr
alloys, addition of Zr is known to have an effect of producing higher
density of \textit{L}1\textsubscript{2} structure precipitation Al\textsubscript{3}Sc\textit{\textsubscript{\textit{x}}}Zr\textsubscript{1-}\textit{\textsubscript{\textit{x}}}
during the coarsening stage\cite{Deschamps20072775,Fuller20055401,Fuller20055415}.
It will be interesting to compare the precipitation kinetics of these
two alloys system. We plot the mean size evolution versus advancement
factor of ternary Fe\textendash{}Cu\textendash{}Ni alloys in Fig.
\ref{fig:Comparison}(a). As for Al\textendash{}Sc\textendash{}Zr
system, the kinetic data are extracted from Ref. \onlinecite{ADEM:ADEM200600246, Clouet-2005}
within the time rage from 0.3s to 0.5s, which seems comparable to
our simulation time range. The cube root of mean size of precipitated
Al\textsubscript{3}Sc\textit{\textsubscript{\textit{x}}}Zr\textsubscript{1-}\textit{\textsubscript{\textit{x}}}
clusters versus total number of atoms in precipitations is represented
in Fig. \ref{fig:Comparison}(b). The difference is quite remarkable,
little change occurred by Ni addition. However, Zr addition significantly
refines Al\textsubscript{3}Sc\textit{\textsubscript{\textit{x}}}Zr\textsubscript{1-}\textit{\textsubscript{\textit{x}}}
precipitation. Clouet \textit{et al.}\cite{Clouet2006} 
revealed that the diffusivity difference of Zr and Sc make Zr-rich external
shell forming around the precipitation, which blocks Ostwald ripening.
Based on the comparison from the figures, Ni plays a different role.
The higher number density in Fe\textendash{}Cu\textendash{}Ni ternary
alloys during coarsening stage is just because Ni somehow slows down
the precipitation of Cu clusters. So Ni indeed has a temporal \textquotedblleft{}delay\textquotedblright{}
effect rather than refinement.

\begin{figure}

\includegraphics[width=8cm]{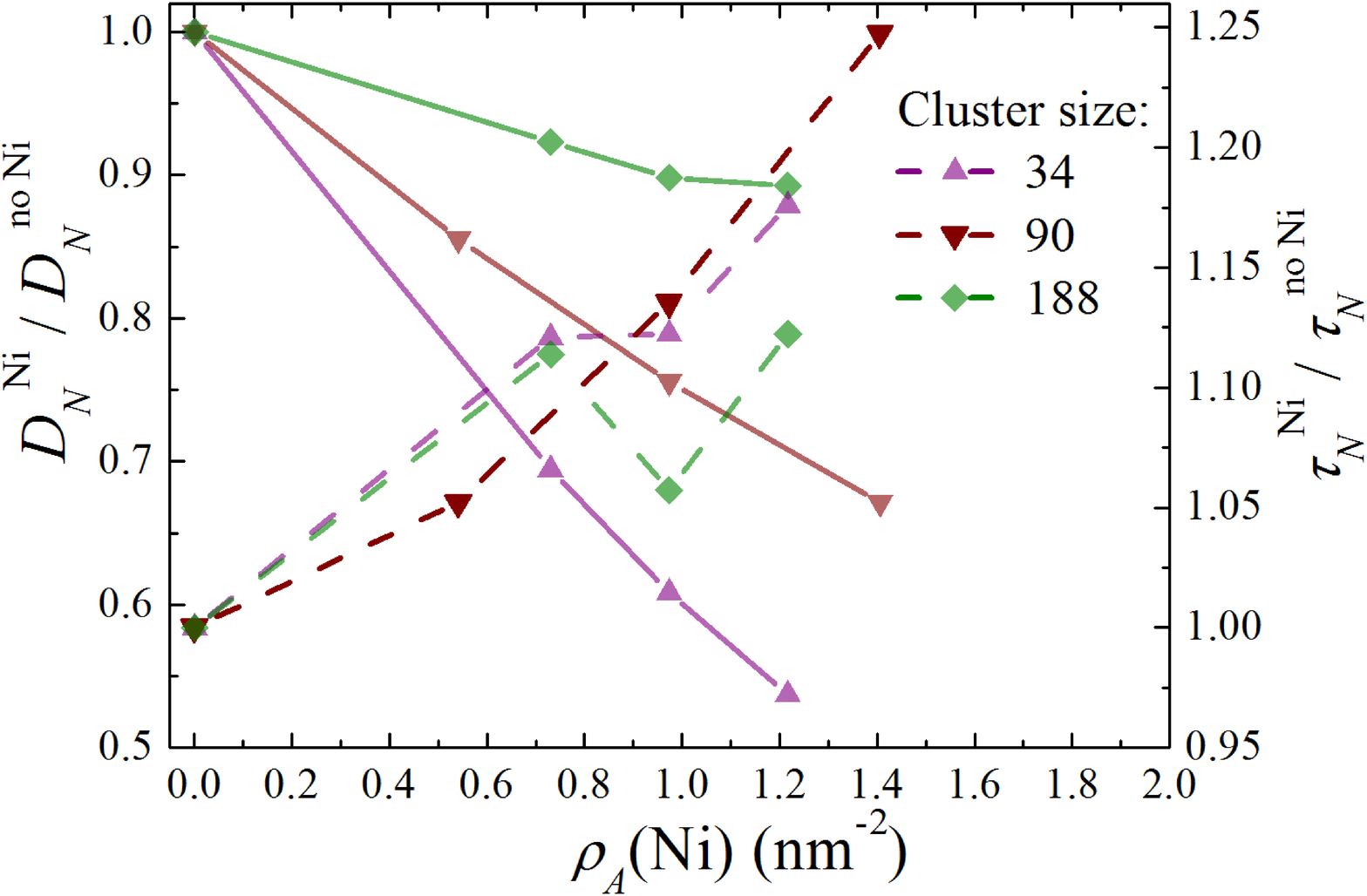}\caption{Cluster mobility dependency of $\rho_{A}\left(\mathrm{Ni}\right)$
at 823K.\label{fig:Cluster-mobility-dependency}}

\end{figure}

It is then natural to apply the cluster mobility analysis
for the ternary alloys. We simulated the mobility of Cu clusters in
Fe\textendash{}Cu\textendash{}Ni alloys at 823K. As is shown in
Fig. \ref{fig:Cluster-mobility-dependency}, diffusion coefficient
decreases linearly with area density of Ni, a minimum of about 50\%
lower diffusion coefficient than binary alloy is found for studied
area densities. While the life time of clusters increases with area density and has
a maximum of about 25\% higher than binary alloy for studied area
densities. So the decreased diffusion coefficient is responsible for
the delay effect of Ni. Even though, the slope of diffusion coefficient
curves deceases with cluster size, larger clusters are less affected
by the area density of Ni. So the delay effect by decreasing diffusion
coefficient will be weaken with the clusters growing larger. Thus
the Cu cluster number density of ternary alloys is expected to converge
to a common value in the long term precipitation, which has been observed
by experiments of Buswell \textit{et al.}\cite{Buswell-1990} and
simulation of Bonny \textit{et al.}\cite{doi:10.1080/14786430903299824}.
Also, the linear law of decreasing diffusion coefficient may explain
the sequence of the delayed curves of advancement factor and number
density of ternary alloys. We actually has an unexpected sequence
of 1.5 at. \% -0.5 at. \% -1.0 at. \% -2.0 at. \%, that is to say Fe\textendash{}1.34 at. \% Cu\textendash{}1.5 at. \% Ni
is evolving faster than other ternary alloys in real time. While MC
time gives an ordered 0.5 at. \% -1.5 at. \% -1.0 at. \% -2.0 at. \% sequence. Since $f_{\mathrm{V}}^{\mathrm{M}}$
has a hyperbolic relationship with Ni content, and diffusion coefficient
has a linear relationship, numerically they will yield a minimum time on a specific Ni content, which means a ternary alloy 
containin this Ni content has fastest evolution. It is possible that this specific Ni content is near 1.5 at. \%.

\section{Conclusion}

We have simulated the precipitation during thermal ageing of binary
Fe\textendash{}Cu and ternary Fe\textendash{}Cu\textendash{}Ni alloys
by AKMC method. The energy model is based on a two-body short range
model. A nonlinear time adjusting method from post-processing data
has been proposed. Using the combined computational techniques, though
the cluster density is over-estimated, good agreement of Cu precipitation
kinetics has been obtained over a temperature range. For the effect
of Ni on Cu precipitation, the following conclusions have been found:
\begin{enumerate}
\item Peak number density of Cu clusters is higher in Fe\textendash{}Cu\textendash{}Ni ternary alloys, which can be
explained by Ni promoting nucleation of Cu clusters. Surface area
density of Ni on clusters has a linear relationship with Ni content;
\item A delay effect has been found for Ni on Cu precipitation in the coarsening
stage. Comparison with Al\textendash{}Sc\textendash{}Zr alloys reveals
this effect is mainly temporal, rather than refinement of precipitations;
\item Diffusion coefficient deceases linearly with area density of Ni, while
life time increases. This Deceasing effect of the cluster diffusion coefficient
 is responsible for the delay effect. 
Even though, this effect weakens with larger cluster size.
\end{enumerate}
On the other side, limitation of this AKMC study still remains. We are working on
 better activation energy model to over-come the shortcoming of pairwise energy model.
\begin{acknowledgments}
This work is funded by the National Natural Science Foundation of China
under grants 51071111 and by the the National High Technology Research and Development Program of China
 (863 Program) under grants 2012AA050901. The authors would like to thank the support by Suzhou Nuclear Power Research Institute.\end{acknowledgments}
\bibliographystyle{apsrev4-1}
\bibliography{kmc}
\end{document}